\documentclass{nature}

\usepackage{lettrine}
\usepackage{graphicx}% Include figure files
\usepackage{dcolumn}% Align table columns on decimal point
\usepackage{bm}% bold math
\usepackage{amssymb}
\usepackage{amssymb}
\usepackage{epsfig}
\usepackage[10pt]{moresize}
\usepackage{comment}
\usepackage{graphicx}
\usepackage{epsfig}
\usepackage{subfigure}

\title{Generation of a macroscopic entangled coherent state using quantum memories in circuit QED}
\author{Tong Liu$^{1}$, Qi-Ping Su$^{1}$, Shao-Jie Xiong$^{1}$, Jin-Ming Liu$^{2}$, Chui-Ping Yang$^{1\star}$, and Franco Nori$^{3,4}$}
\begin{document}

\maketitle
\begin{affiliations}
\item Department of Physics, Hangzhou Normal University,
Hangzhou, Zhejiang 310036, China
\item State Key Laboratory of Precision Spectroscopy,
Department of Physics, East China Normal University, Shanghai
200062, China
\item CEMS, RIKEN, Saitama 351-0198, Japan
\item Department of Physics, The University of Michigan, Ann Arbor, Michigan 48109-1040, USA
$^\star$ Email: yangcp@hznu.edu.cn
\end{affiliations}
%\date{\today}

\begin{abstract}
$W$-type entangled states can be used as quantum channels for, e.g., quantum
teleportation, quantum dense coding, and quantum key distribution. In
this work, we propose a way to generate a macroscopic $W$-type entangled
coherent state using quantum memories in circuit QED. The memories considered
here are nitrogen-vacancy center ensembles (NVEs), each located in a
different cavity. This proposal does not require initially preparing each
NVE in a coherent state instead of a ground state, which should significantly
reduce its experimental difficulty. For most of the operation time, each
cavity remains in a vacuum state, thus decoherence caused by the cavity
decay and the unwanted inter-cavity crosstalk are greatly suppressed. Moreover,
only one external-cavity coupler qubit is needed. This method is quite general and can be applied to
generate the proposed $W$ state with atomic ensembles or other spin
ensembles distributed in different cavities.
\end{abstract}

\lettrine[lines=2]{U}nlike bipartite systems, it has been proven that there
exist two inequivalent classes of multipartite entangled states, such as GHZ
states [1] and $W$ states [2], which cannot be converted to each other by
local operations and classical communications. Relative to the tripartite
entangled states, GHZ states are fragile: if any one qubit is traced out,
the remaining bipartite states are separable states. However, $W$ states are
robust against qubit loss and qubit-flip noise because they maintain
bipartite entanglement. $W$ states are important for quantum communications.
For example, $W$ states can be used as quantum channels for quantum
teleportation [3], quantum dense coding [4], and quantum key distribution
[5].

Over the past years, a number of theoretical ideas have been proposed for
creating a \textit{discrete-variable} $W$-class entangled state $%
|W_{n-1,1}\rangle _{DV}=\frac{1}{\sqrt{n}}\sum \ P_{z}|0\rangle ^{\otimes {%
(n-1)}}|1\rangle $ of qubits (i.e., \textit{two-state} particles or \textit{%
two-level} quantum systems) [6-13], where $P_{z}$ is the symmetry
permutation operator for the qubits ($1,2\cdots n$), and $\sum \
P_{z}|0\rangle ^{\otimes {(n-1)}}|1\rangle $ denotes the totally-symmetric
state in which ($n-1$) qubits out of a total of $n$ qubits are in the state $%
|0\rangle $, while the remaining qubit is in the state $|1\rangle $. As an
example, consider a three-qubit case (i.e., $n=3$), for which the $W$ state
is $\left\vert W_{2,1}\right\rangle _{DV}=\frac{1}{\sqrt{3}}\left(
\left\vert 001\right\rangle +\left\vert 010\right\rangle +\left\vert
100\right\rangle \right) $. Experimentally, the discrete-variable $W$ states
$|W_{n-1,1}\rangle _{DV}$ have been created with up to eight trapped ions
[14], four optical modes [15], three superconducting phase qubits coupled
capacitively [16], atomic ensembles in four quantum memories [17], and two
superconducting phase qubits plus a resonant cavity [18].

On the other hand, there is much interest in \textit{entangled coherent
states} (ECSs) [19-28]. In this work we focus on a macroscopic $W$-type ECS
(i.e., \textit{continuous-variable} $W$ state), described by
\begin{eqnarray}
|W_{n-1,1}\rangle _{CV} &=&c_{0}|-\alpha \rangle |\alpha \rangle ...|\alpha
\rangle +c_{1}|\alpha \rangle |-\alpha \rangle |\alpha \rangle ...|\alpha
\rangle +...  \nonumber \\
&&+c_{n-1}|\alpha \rangle ...|\alpha \rangle |-\alpha \rangle ,
\end{eqnarray}%
where $\sum_{i=0}^{n-1}\left\vert c_{i}\right\vert ^{2}=1$, with $c_{i}\neq 0
$ ($i=0,1,...,n-1$), $\left\vert \alpha \right\rangle $ ($\left\vert -\alpha
\right\rangle $) is a coherent state, $\alpha $ is a complex number, and $%
\left\langle \alpha \right\vert \left. -\alpha \right\rangle =\exp \left(
-2\left\vert \alpha \right\vert ^{2}\right) \simeq 0$, when $\left\vert
\alpha \right\vert $ is large enough. The $W$ state (1) is of fundamental
interest in quantum mechanics and plays an important role in quantum
information processing (QIP) and quantum communications. For instance, the $W
$ state (1) can be used to test quantum nonlocality without inequality
[29,30] and the violation of the Bell inequalities because such state is
greater than that for any states involving two spin-1/2 particles [30,31].
In addition, Ref.~[32] has shown that there exists a quantum information
protocol which is not suitable for GHZ-type ECSs but can only be
accomplished with the $W$ state (1). Moreover, the $W$ state (1) is a
necessary resource for remote symmetric entanglement [32], which allows two
distant parties to share a symmetric entangled state. For the past years,
theoretical methods have been proposed for generating the $W$ state (1) in
some physical systems [33-37]. Refs.~[32-34] have proposed how to generate
the $W$ state (1) of three/four modes with linear optical devices, and Refs.
[36,37] have discussed how to create the $W$ state (1) of three-cavity
fields based on cavity QED. However, in these schemes, the $W$ ECSs were
prepared with photons or cavity fields, and thus decoherence may pose a
problem due to photon loss or cavity-field decay.

Hybrid quantum systems, composed of superconducting qubits, nitrogen-vacancy
centers (NVCs), nitrogen-vacancy center ensembles (NVEs), or/and
superconducting microwave resonators/cavities, have attracted tremendous
attention [38-41]. Recently, much progress has been made in this field. For
instance, coherent coupling between a superconducting flux/transmon qubit
and an NVE [42,43] or between an NVC/NVE and a superconducting resonator
[44,45] has been experimentally demonstrated. Moreover, based on the hybrid
systems, various quantum operations, such as entanglement preparation,
quantum logic gates, and information transfer, have been investigated in
theory [40,46-49] and demonstrated in experiment [42,50,51].

Inspired by previous works and the long decoherence time of NVEs, we here
consider a hybrid system composed of one-dimensional transmission line
resonators (TLRs) each hosting an NVE and a qubit and connected to a coupler
qubit $A$ [Fig.~1(a), Fig.~2]. We then propose a way to generate a
continuous-variable $W$-type entangled coherent state, described by Eq.~(1),
by using NVEs each located in a different cavity. Because of the long
decoherence time of NVEs, the prepared $W$ state can be stored for a long
time. Note that NVEs have been recently considered as good memory elements
in quantum information processing [39,40,42,45-49,51].

As shown below, this proposal has the following features and advantages: (i)
Different from the previous works [33-37], the $W$ state is prepared using
NVEs (quantum memories) \textit{instead of cavity photons}. Thus, the
prepared $W$ state can be stored for a long time due to the long decoherence
time of the NVEs. (ii) Because cavity photons are virtually excited for most
of the operation time, decoherence caused by the cavity decay and the
unwanted inter-cavity cross talk is greatly suppressed. (iii) Each NVE is
initially in the ground state. Thus, there is no need to initially prepare
each NVE in a coherent state, which should greatly reduce its experimental
difficulty. (iv) Moreover, only one external-cavity coupler qubit is needed,
which simplifies the circuit. This method is quite general and can be
applied to prepare the proposed $W$ state with atomic ensembles or other
spin ensembles based on cavity/circuit QED.

There are several additional motivations of this proposal:

(i) Planar superconducting TLRs with internal quality factors above one
million ($Q>10^{6}$) have been recently reported [52], for which the
lifetime of microwave photons can reach $\sim 1$ ms. Comparably, a lifetime
of $\sim 1$ s for an NVE has been experimentally reported [53]. Hence, a NVE
is a good memory element for storing quantum states, superior to using
cavity photons as memories.

(ii) By location operations, the prepared $W$ state of the NVEs can be
mapped onto the cavities (see the \textquotedblleft Quantum state
transfer\textquotedblright\ subsection).

(iii) The NVEs could be prepared in the ground state at a 40$-$50 mK or
higher temperature [42,44]. The strong coupling of a superconducting qubit
with a microwave resonator (e.g., $g/2\pi \sim 360$ MHz for a transmon qubit
coupled to a TLR [54,55]) has been reported in experiments, and the strong
coupling ($\sim 11$ MHz) of an NVE to a TLR has recently been experimentally
demonstrated [44]. Moreover, superconducting qubits, capacitively or
inductively coupled to TLRs [13,56-68], were previously employed for QIP.
Hence, the model considered in this work is reasonable and physical.

Note that based on circuit QED, a number of proposals have been presented
for creating entangled states (e.g., Bell states, NOON states, and GHZ
states)\ of \textit{microwave photons} distributed in different
TLRs/cavities [57,58,60,63,65,67]. Instead of preparing entangled states of
cavity microwave photons, this work focuses on preparing the NVEs in a
\textit{continuous-variable} $W$-type entangled coherent state.

In this work we will also discuss possible experimental implementation of
our proposal and numerically calculate the operational fidelity for
generating a $W$-type entangled coherent state of three NVEs. Our numerical
simulation shows that highly-fidelity implementation of $W$-type entangled
coherent states with three NVEs is feasible with current circuit QED
technology. The numerical calculations in this work were performed using the
QuTiP software [69,70].

\section*{Results}

\textbf{$W$-state preparation.} Consider a hybrid system consisting of a
coupler qubit $A$ and three cavities, each hosting a qubit and an NVE
[Fig.~1(a)]. Each cavity here is a one-dimensional transmission line
resonator. The qubit and the NVE placed in cavity $j$ are labelled as qubit $%
j$ and NVE $j$ ($j=1,2,3$). The two levels of qubit $A$ are denoted as $%
|g\rangle _{A}$ and $|e\rangle _{A},$ while those of qubit $j$ as $|g\rangle
_{j}$ and $|e\rangle _{j}$. The coupling and decoupling of each qubit from
its cavity (cavities) can be achieved by prior adjustment of the qubit level
spacings or the cavity frequency. For superconducting devices, their level
spacings can be rapidly (within $1$--$3$ ns [65,71,72]) adjusted by varying
external control parameters (e.g., via changing the external magnetic flux
threading the superconducting loop of phase, transmon, Xmon or flux qubits;
see, e.g., [71-80]). In addition, as described in the Methods section, the
coupling and decoupling of an NVE with a cavity can be made by rapidly
adjusting the cavity frequency [81,82].

Assume that the qubits, cavities, and NVEs are initially decoupled from one
another [Fig.~1(b)]. The procedure for generating a $W$-type entangled
coherent state of the three NVEs is described below:

\textbf{Step 1.} Adjust the level spacings of the coupler qubit $A$ so that
it is resonantly coupled to each cavity [Fig.~1(c)]. Assume that the
coupling constant of qubit $A$ with cavity $j$ is $g_{A_{j}}$. In the
interaction picture, the Hamiltonian reads
\begin{equation}
H_{I1}=\sum\limits_{j=1}^{3}g_{A_{j}}(a_{j}^{\dagger }\sigma
_{A}^{-}+a_{j}\sigma _{A}^{+}),
\end{equation}%
where $\sigma _{A}^{+}=|e\rangle _{A}\langle g|$ and $\sigma
_{A}^{-}=|g\rangle _{A}\langle e|$ are the raising and lowering operators
for qubit $A$, while $a_{j}$ and $a_{j}^{\dagger }$ are the annihilation and
creation operators for the mode of cavity $j$ $(j=1,2,3)$. We set $%
g_{A_{1}}=g_{A_{2}}=g_{A_{3}}=g_{A}$, which can be met by a prior design of
the sample with appropriate values of the coupling capacitance $C_{1},$ $%
C_{2},$ and $C_{3}$. Assume now that qubit $A$ is initially in the state $%
|e\rangle _{A}$ and each cavity is initially in the vacuum state. It is easy
to show that the state $\prod\limits_{j=1}^{3}|0\rangle _{c_{j}}\otimes
|e\rangle _{A}$ of the system, under the Hamiltonian (2), evolves into
\begin{equation}
\cos (\sqrt{3}g_{A}t)\prod\limits_{j=1}^{3}|0\rangle _{c_{j}}\otimes
|e\rangle _{A}-i\sin (\sqrt{3}g_{A}t)|W_{2,1}\rangle _{c}\otimes |g\rangle
_{A}.
\end{equation}%
Here, the state $|W_{2,1}\rangle _{c}$ of the three cavities (1,2,3) is
given by
\begin{equation}
|W_{2,1}\rangle _{c}=\frac{1}{\sqrt{3}}(|1\rangle |0\rangle |0\rangle
+|0\rangle |1\rangle |0\rangle +|0\rangle |0\rangle |1\rangle ),
\end{equation}%
where $|i\rangle |j\rangle |k\rangle $ is the abbreviation of the state $%
|i\rangle _{c_{1}}$$|j\rangle _{c_{2}}$$|k\rangle _{c_{3}}$ of cavities
(1,2,3) with $i,j,k\in \{0,1\}$; $|0\rangle $ and $|1\rangle $ represent the
vacuum state and the single-photon state, respectively. From Eq. (3), it can
be seen that when the interaction time equals to $t=\pi /\left( 2\sqrt{3}%
g_{A}\right) $, we can create the state $|W_{2,1}\rangle _{c}$ of the three
cavities (1,2,3). Note that the coupler qubit $A$ is in the ground state $%
|g\rangle _{A}$ after the operation here and will remain in the ground state
$|g\rangle _{A}$ during the rest of the operations below.

\textbf{Step 2.} Adjust the level spacings of qubit $A$ back to the original
level structure such that it is decoupled from each cavity. In addition,
adjust the level spacing of intra-cavity qubit $j$ such that qubit $j$ is
resonantly coupled to cavity $j$ [Fig.~1(d)]. The resonant coupling constant
of qubit $j$ with cavity $j$ is denoted as $g_{rj}$. In the interaction
picture, the Hamiltonian can be written as
\begin{equation}
H_{I2}=\sum\limits_{j=1}^{3}g_{rj}(a_{j}^{\dagger }\sigma
_{j}^{-}+a_{j}\sigma _{j}^{+})
\end{equation}%
where $\sigma _{j}^{+}=|e\rangle _{j}\langle g|$ and $\sigma
_{j}^{-}=|g\rangle _{j}\langle e|$ are the raising and lowering operators
for qubit $j$. For simplicity, we set $g_{r1}=g_{r2}=g_{r3}=g_{r},$ which
can be achieved by tuning the level spacings of qubit $j$ or adjusting the
position of qubit $j$ in cavity $j$ ($j=1,2,3$). It is easy to show that
under this Hamiltonian (5), the time evolution of the state $|g\rangle
_{j}|n\rangle _{c_{j}}$ of qubit $j$ and cavity $j$ is described by
\begin{equation}
|g\rangle _{j}|n\rangle _{c_{j}}\rightarrow \cos (\sqrt{n}g_{r}t)|g\rangle
_{j}|n\rangle _{c_{j}}-i\sin (\sqrt{n}g_{r}t)|e\rangle _{j}|n-1\rangle
_{c_{j}},
\end{equation}%
where $|n\rangle _{c_{j}}$ and $|n-1\rangle _{c_{j}}$ are the photon-number
states of cavity $j$. Assume now that qubit $j$ is initially in the state $%
|g\rangle _{j}$. Choosing $t=\pi /\left( 2g_{r}\right) $, one obtains the
transformation $|g\rangle _{j}|1\rangle _{c_{j}}\rightarrow -i|e\rangle
_{j}|0\rangle _{c_{j}}$. As a result, the state $|W_{2,1}\rangle _{c}$ of
the three cavities turns into the following state of the three intracavity
qubits (1,2,3)
\begin{equation}
|W_{2,1}\rangle =\frac{1}{\sqrt{3}}(|e\rangle |g\rangle |g\rangle +|g\rangle
|e\rangle |g\rangle +|g\rangle |g\rangle |e\rangle ),
\end{equation}%
where $|i\rangle |j\rangle |k\rangle $ is the abbreviation of the state $%
|i\rangle _{1}$$|j\rangle _{2}$$|k\rangle _{3}$ of intracavity qubits
(1,2,3) with $i,j,k\in \{g,e\}$. It should be noted that each cavity returns
to its original vacuum state after the operation here and will remain in the
vacuum state during the following operations.

The condition $g_{r1}=g_{r2}=g_{r3}=g_{r}$ is unnecessary. For the case of $%
g_{r1}\neq g_{r2}\neq g_{r3},$ one can still obtain the state (7) from the
state (4), by adjusting the level spacings of qubit $j$ to bring qubit $j$
on resonance with cavity $j$ for a time $t_{j}=$ $\pi /\left( 2g_{rj}\right)
$ ($j=1,2,3$).

\textbf{Step 3.} Adjust the level spacings of intracavity qubits back to the
original level configuration, such that they are decoupled from their
cavities. Then apply a classical pulse to qubit $j.$ The pulse is resonant
with the $|g\rangle _{j}\leftrightarrow |e\rangle _{j}$ transition of qubit $%
j$ [Fig.~1(e)]. The interaction Hamiltonian in the interaction picture is
given by
\begin{equation}
H_{I3}=\sum\limits_{j=1}^{3}\Omega _{eg_{j}}(e^{i\phi }|g\rangle _{j}\langle
e|+h.c.),
\end{equation}%
where $\Omega _{eg_{j}}$ and $\phi $ are the Rabi frequency and the initial
phase of the pulse, respectively. Set $\Omega _{eg_{1}}=\Omega
_{eg_{2}}=\Omega _{eg_{3}}=\Omega _{eg},$ which can be readily met by
adjusting the pulse intensities. It is easy to find that under the
Hamiltonian (8), one can obtain the following rotations
\begin{eqnarray}
|g\rangle _{j} &\rightarrow &\cos (\Omega _{eg}t)|g\rangle _{j}-ie^{-i\phi
}\sin (\Omega _{eg}t)|e\rangle _{j},  \nonumber \\
|e\rangle _{j} &\rightarrow &\cos (\Omega _{eg}t)|e\rangle _{j}-ie^{i\phi
}\sin (\Omega _{eg}t)|g\rangle _{j}.
\end{eqnarray}%
We set $t=\pi /\left( 4\Omega _{eg}\right) $ and $\phi =-\pi /2$ to pump the
state $|e\rangle _{j}$ to $|-\rangle _{j}$ and $|g\rangle _{j}$ to $%
|+\rangle _{j}$. Here, $|\pm \rangle _{j}=(|e\rangle _{j}\pm |g\rangle _{j})/%
\sqrt{2}$ are the rotated basis states of qubit $j$. Thus, the state (7)
becomes
\begin{equation}
|\widetilde{W}_{2,1}\rangle =\frac{1}{\sqrt{3}}(|-\rangle |+\rangle
|+\rangle +|+\rangle |-\rangle |+\rangle +|+\rangle |+\rangle |-\rangle ).
\end{equation}

\textbf{Step 4.} Adjust the frequency of each cavity such that cavity $j$
interacts with qubit $j$ and NVE $j$ [Fig.~1(f)]. Then apply a classical
pulse (with frequency $\omega _{j}$ equal to $\omega _{eg_{j}}$) to qubit $j$
[Fig.~1(f)]. Here, $\omega _{eg_{j}}$ is the $|g\rangle \leftrightarrow
|e\rangle $ transition frequency of qubit $j$. The system Hamiltonian in the
interaction picture yields
\begin{eqnarray}
H_{I4} &=&\sum\limits_{j=1}^{3}g_{j}\left[ \exp \left( i\delta
_{a_{j}}t\right) a_{j}^{\dagger }\sigma _{j}^{-}+h.c.\right]
+\sum\limits_{j=1}^{3}g_{b_{j}}\left[ \exp \left( i\delta _{b_{j}}t\right)
a_{j}^{\dagger }b_{j}+h.c.\right]  \nonumber \\
&&+\sum\limits_{j=1}^{3}\Omega _{j}(\sigma _{j}^{+}+\sigma _{j}^{-}),
\end{eqnarray}%
where $\delta _{a_{j}}=\omega _{c_{j}}-\omega _{eg_{j}}$ and $\delta
_{b_{j}}=\omega _{c_{j}}-\omega _{b_{j}}$ are the frequency detunings ($%
\omega _{c_{j}}$ being the frequency of cavity $j$ while $\omega _{b_{j}}$
being the frequency of a bosonic mode describing NVE $j$), $b_{j}$ is the
bosonic operator for NVE $j$, $g_{j}$ is the off-resonant coupling constant
of qubit $j$ with cavity $j$, $g_{b_{j}}$ is the coupling constant of NVE $j$
with cavity $j$, and $\Omega _{j}$ is the Rabi frequency of the pulse
applied to qubit $j$ [Fig. 1(f)]. Note that the second term of Eq.~(11)
describes three NVEs interacting with their respective cavities (see the
Methods section). In a rotated basis \{$|+\rangle _{j},|-\rangle _{j}$\},
one has $\sigma _{j}^{+}=\left( \widetilde{\sigma }_{z_{j}}-\widetilde{%
\sigma }_{j}^{+}+\widetilde{\sigma }_{j}^{-}\right) /2$ and $\sigma
_{j}^{-}=\left( \widetilde{\sigma }_{z_{j}}+\widetilde{\sigma }_{j}^{+}-%
\widetilde{\sigma }_{j}^{-}\right) /2$, where $\widetilde{\sigma }%
_{z_{j}}=|+\rangle _{j}\langle +|-|-\rangle _{j}\langle -|$ , $\widetilde{%
\sigma }_{j}^{+}=|+\rangle _{j}\langle -|,$ and $\widetilde{\sigma }%
_{j}^{-}=|-\rangle _{j}\langle +|$. Hence, the Hamiltonian (11) can be
expressed as
\begin{eqnarray}
H_{I4} &=&\sum\limits_{j=1}^{3}\frac{1}{2}g_{j}\left[ \exp \left( i\delta
_{a_{j}}t\right) a_{j}^{\dagger }(\widetilde{\sigma }_{z_{j}}+\widetilde{%
\sigma }_{j}^{+}-\widetilde{\sigma }_{j}^{-})+h.c.\right]  \nonumber \\
&&+\sum\limits_{j=1}^{3}g_{b_{j}}\left[ \exp \left( i\delta _{b_{j}}t\right)
a_{j}^{\dagger }b_{j}+h.c.\right] +\sum\limits_{j=1}^{3}\Omega _{j}%
\widetilde{\sigma }_{z_{j}}.
\end{eqnarray}%
In a new interaction picture under the Hamiltonian $H_{0}^{\prime
}=\sum\limits_{j=1}^{3}\Omega _{j}\widetilde{\sigma }_{z_{j}}$, one obtains
from Eq.~(12)
\begin{eqnarray}
H_{I4} &=&\sum\limits_{j=1}^{3}\frac{1}{2}g_{j}\left[ \exp \left( i\delta
_{a_{j}}t\right) a_{j}^{\dagger }(\widetilde{\sigma }_{z_{j}}+e^{2i\Omega
_{j}t}\widetilde{\sigma }_{j}^{+}-e^{-2i\Omega _{j}t}\widetilde{\sigma }%
_{j}^{-})+h.c.\right]  \nonumber \\
&&+\sum\limits_{j=1}^{3}g_{b_{j}}\left[ \exp \left( i\delta _{b_{j}}t\right)
a_{j}^{\dagger }b_{j}+h.c.\right] .
\end{eqnarray}%
In the strong-driving regime $2\Omega _{j}\gg \{g_{j},\delta _{a_{j}}\}$,
one can apply a rotating-wave approximation and eliminate the terms that
oscillate with high frequencies. Thus, the Hamiltonian~(13) becomes
\begin{equation}
H_{I4}=\sum\limits_{j=1}^{3}\frac{1}{2}g_{j}\widetilde{\sigma }_{z_{j}}\left[
\exp \left( i\delta _{a_{j}}t\right) a_{j}^{\dagger }+h.c.\right]
+\sum\limits_{j=1}^{3}g_{b_{j}}\left[ \exp \left( i\delta _{b_{j}}t\right)
a_{j}^{\dagger }b_{j}+h.c.\right] .
\end{equation}%
Consider now the large detuning conditions $\delta _{a_{j}}\gg g_{j}$ and $%
\delta _{b_{j}}\gg g_{b_{j}}.$ It is straightforward to show that the
Hamiltonian (14) changes to (for details, see Ref.~[83])
\begin{eqnarray}
H_{\mathrm{eff}} &=&\sum\limits_{j=1}^{3}\frac{g_{b_{j}}^{2}}{\delta _{b_{j}}%
}(b_{j}b_{j}^{\dagger }a_{j}^{\dagger }a_{j}-a_{j}a_{j}^{\dagger
}b_{j}^{\dagger }b_{j})  \nonumber \\
&-&\sum\limits_{j=1}^{3}\lambda _{j}\widetilde{\sigma }_{z_{j}}\left[ \exp
\left( -i\delta _{c_{j}}t\right) b_{j}+\exp \left( i\delta _{c_{j}}t\right)
b_{j}^{\dagger }\right] ,
\end{eqnarray}%
where $\lambda _{j}=\frac{g_{j}g_{b_{j}}}{4}(1/\delta _{a_{j}}+1/\delta
_{b_{j}})$ and $\delta _{c_{j}}=\delta _{a_{j}}-\delta _{b_{j}}$. As
mentioned previously, each cavity is in the vacuum state after the first
three steps of operation above. In this case, the Hamiltonian (15) reduces
to
\begin{equation}
H_{\mathrm{eff}}=-\sum\limits_{j=1}^{3}\frac{g_{b_{j}}^{2}}{\delta _{b_{j}}}%
b_{j}^{\dagger }b_{j}-\sum\limits_{j=1}^{3}\lambda _{j}\widetilde{\sigma }%
_{z_{j}}\left[ \exp \left( -i\delta _{c_{j}}t\right) b_{j}+\exp \left(
i\delta _{c_{j}}t\right) b_{j}^{\dagger }\right] ,
\end{equation}%
where the first term is the vacuum contribution Stark shift of NVEs, while
the second term describes the coupling between qubit $j$ and NVE $j,$
mediated by the mode of cavity $j$. Because of using the large detuning
technique, the effective coupling $\lambda _{j}$ is smaller than $g_{j}$ or $%
g_{b_{j}}$ by at least one order of magnitude. Accordingly, the operation
time for this last step of the operation (essentially based on a model via
virtual transitions) would become longer by one order of magnitude, when
compared with each of the first three steps of operation via resonant
interaction.

In a new interaction picture under the Hamiltonian $H_{0}^{^{\prime \prime
}}=-\sum\limits_{j=1}^{3}\frac{g_{b_{j}}^{2}}{\delta _{b_{j}}}b_{j}^{\dagger
}b_{j}$, the effective Hamiltonian (16) can be rewritten as
\begin{equation}
H_{\mathrm{eff}}=-\sum\limits_{j=1}^{3}\lambda _{j}\widetilde{\sigma }%
_{z_{j}}(b_{j}e^{-i\Delta _{j}t}+b_{j}^{\dagger }e^{i\Delta _{j}t}),
\end{equation}%
where $\Delta _{j}=\delta _{c_{j}}-g_{b_{j}}^{2}/\delta _{b_{j}}$.

Let us now assume that the NVEs are initially in the state $%
\prod\limits_{j=1}^{3}|0\rangle _{b_{j}}$. Thus, under the Hamiltonian~(17),
the joint state $|\widetilde{W}_{2,1}\rangle \otimes
\prod\limits_{j=1}^{3}|0\rangle _{b_{j}}$ of the three intracavity qubits
and the three NVEs evolves into
\begin{eqnarray}
\frac{1}{\sqrt{3}}(|-\rangle |+\rangle |+\rangle |-\alpha \rangle |\alpha
\rangle |\alpha \rangle +|+\rangle |-\rangle |+\rangle |\alpha \rangle
|-\alpha \rangle |\alpha \rangle +|+\rangle |+\rangle |-\rangle |\alpha
\rangle |\alpha \rangle |-\alpha \rangle ),
\end{eqnarray}
with
\begin{equation}
\alpha _{j}=\frac{\lambda _{j}}{\Delta _{j}}(e^{i\Delta _{j}t}-1).
\end{equation}%
Here, $\left\vert \alpha \right\rangle $ ($\left\vert -\alpha \right\rangle $%
) is a coherent state and we have set $\alpha _{1}=\alpha _{2}=\alpha
_{3}=\alpha $ for simplicity (which can be met for identical qubits, NVEs,
and cavities). After returning to the original interaction picture by
performing a unitary transformation $U=e^{-iH_{0}^{^{\prime
}}t}e^{-iH_{0}^{^{\prime \prime }}t}$, the state (18) becomes
\begin{eqnarray}
\left\vert \varphi \right\rangle &=&\frac{1}{\sqrt{3}}(|-\rangle |+\rangle
|+\rangle |-\beta \rangle |\beta \rangle |\beta \rangle +|+\rangle |-\rangle
|+\rangle |\beta \rangle |-\beta \rangle |\beta \rangle  \nonumber \\
&&+|+\rangle |+\rangle |-\rangle |\beta \rangle |\beta \rangle |-\beta
\rangle ),
\end{eqnarray}%
where a common phase factor is discarded, $\left\vert \beta \right\rangle $ (%
$\left\vert -\beta \right\rangle $) is a coherent state, and
\begin{equation}
\beta =\alpha e^{ig_{b_{1}}^{2}t/\delta _{b_{1}}}=\alpha
e^{ig_{b_{2}}^{2}t/\delta _{b_{2}}}=\alpha e^{ig_{b_{3}}^{2}t/\delta
_{b_{3}}}
\end{equation}%
for
\begin{equation}
g_{b_{1}}^{2}/\delta _{b_{1}}=g_{b_{2}}^{2}/\delta
_{b_{2}}=g_{b_{2}}^{2}/\delta _{b_{2}}.
\end{equation}%
The condition~(21) is automatically satisfied for identical NVEs and
cavities. The state~(20) can be expressed as
\begin{eqnarray}
\left\vert \varphi \right\rangle =\frac{1}{2\sqrt{2}} &[&|W_{1}\rangle
(|e\rangle |e\rangle |e\rangle -|g\rangle |g\rangle |g\rangle
)+|W_{2}\rangle (|e\rangle |e\rangle |g\rangle -|g\rangle |g\rangle
|e\rangle )  \nonumber \\
&&+|W_{3}\rangle (|e\rangle |g\rangle |e\rangle -|g\rangle |e\rangle
|g\rangle )+|W_{4}\rangle (|e\rangle |g\rangle |g\rangle -|g\rangle
|e\rangle |e\rangle )~],
\end{eqnarray}%
where $|W_{1}\rangle $, $|W_{2}\rangle $, $|W_{3}\rangle $ and $%
|W_{4}\rangle $ are the macroscopic $W$-type entangled coherent states of
three NVEs, given by
\begin{eqnarray}
|W_{1}\rangle &=&\frac{1}{\sqrt{3}}\left( |-\beta \rangle |\beta \rangle
|\beta \rangle +|\beta \rangle |-\beta \rangle |\beta \rangle +|\beta
\rangle |\beta \rangle |-\beta \rangle \right) ,  \nonumber \\
|W_{2}\rangle &=&\frac{1}{\sqrt{3}}\left( |-\beta \rangle |\beta \rangle
|\beta \rangle +|\beta \rangle |-\beta \rangle |\beta \rangle -|\beta
\rangle |\beta \rangle |-\beta \rangle \right) ,  \nonumber \\
|W_{3}\rangle &=&\frac{1}{\sqrt{3}}\left( |-\beta \rangle |\beta \rangle
|\beta \rangle -|\beta \rangle |-\beta \rangle |\beta \rangle +|\beta
\rangle |\beta \rangle |-\beta \rangle \right) ,  \nonumber \\
|W_{4}\rangle &=&\frac{1}{\sqrt{3}}\left( |-\beta \rangle |\beta \rangle
|\beta \rangle -|\beta \rangle |-\beta \rangle |\beta \rangle -|\beta
\rangle |\beta \rangle |-\beta \rangle \right) .
\end{eqnarray}

Now a measurement is separately performed on each intra-cavity qubit along a
measurement basis \{$\left\vert g\right\rangle ,\left\vert e\right\rangle $%
\}. If qubits (1,2,3) are measured in the state (i) $|e\rangle |e\rangle
|e\rangle $ or $|g\rangle |g\rangle |g\rangle $, (ii) $|e\rangle |e\rangle
|g\rangle $ or $|g\rangle |g\rangle |e\rangle $, (iii) $|e\rangle |g\rangle
|g\rangle $ or $|g\rangle |e\rangle |e\rangle $, and (iv) $|e\rangle
|g\rangle |g\rangle $ or $|g\rangle |e\rangle |e\rangle $, one can see from
Eq.~(23) that the three NVEs are respectively prepared in the $W$ states $%
|W_{1}\rangle $, $|W_{2}\rangle $, $|W_{3}\rangle $ and $|W_{4}\rangle $,
respectively.

This method can be extended to a more general case. Consider a hybrid system
composed of $n$ cavities, each hosting a qubit $j$ and an NVE $j$ $%
(j=1,2\cdots n)$ and connected to a coulper qubit $A$, as shown in Fig.~2.
Assume that the initial state of the system is $\prod\limits_{j=1}^{n}|0%
\rangle _{c_{j}}\otimes |e\rangle _{A}\otimes
\prod\limits_{j=1}^{n}|g\rangle _{j}\otimes \prod\limits_{j=1}^{n}|0\rangle
_{b_{j}}$. Employing the four-step procedure described above, it is
straightforward to show that the $n$ NVEs can be prepared in a $W$-type
entangled coherent state. Let $m_{j}=0$ represent qubit $j$ being measured
in the state $|g\rangle $, while $m_{j}=1$ indicates qubit $j$ being
measured in the state $|e\rangle $. If the $n$ intracavity qubits are
measured in the state $|m_{1}m_{2}\cdots m_{n}\rangle $, the $n$ NVEs will
be prepared in the macroscopic $W$-type entangled coherent state
\begin{eqnarray}
\frac{1}{\sqrt{n}} &[&(-1)^{m_{1}}|-\beta \rangle |\beta \rangle |\beta
\rangle \cdots |\beta \rangle +(-1)^{m_{2}}|\beta \rangle |-\beta \rangle
|\beta \rangle \cdots |\beta \rangle  \nonumber \\
&+&\cdots +(-1)^{m_{n}}|\beta \rangle |\beta \rangle |\beta \rangle \cdots
|-\beta \rangle ~].
\end{eqnarray}

Before ending this section, several points need to be addressed as follows:

(i) From the description given above, one can see that only resonant
interactions are used for the first three steps of operation, which can thus
be completed within a very short time (e.g., by increasing the pulse Rabi
frequencies and the qubit-cavity coupling constants). In contrast, the last
step of operation employs a large detuning, leading to a relatively long
operation time. However, cavity photons were virtually excited during this
step of operation. Hence, in the present proposal each cavity remains in a
vacuum state for most of the operation time.

(ii) The adjustment of the cavity frequency during the last step of
operation is unnecessary. Alternatively, one can adjust the level spacings
of the NVEs (by varying the external magnetic fields applied to the NVEs
[48,84]), such that the cavities are coupled with the NVEs or decoupled from
the NVEs.

(iii) As shown above, the intracavity-qubit $W$ state of Eq.~(7) can be
produced within a very short time, because the first two steps of operation,
for producing this intracavity-qubit $W$ state (7), employ resonant
interactions. Alternatively, this intracavity-qubit $W$ state (7) can be
prepared via a detuned interaction between the coupler qubit $A$ and each
cavity [13,64,68]. Thus, there are no cavity photons excited during the
entire state preparation. However, the time required for preparing the $W$
state (7) becomes much longer due to the use of a detuned interaction, and
thus decoherence from the qubits may pose a significant problem.

(iv) Placing a qubit in each cavity [Fig.~1(a)] is necessary in view of
energy conservation. During the last step, each cavity remains in a vacuum
state and thus there is no energy transfer from each cavity onto the NVEs.
Note that the intracavity qubits are the ones that absorb energy from the
pulses applied to them and then transfer their energy to the NVEs through
interaction with the NVEs. Thus, in spite of initially being in the ground
state, the NVEs can be prepared in a $W$-type entangled coherent state.

(v) As discussed previously, a measurement of the states of each
intra-cavity qubit is needed during preparation of the $W$-class entangled
coherent states. To the best of our knowledge, all existing proposals for
creating entangled coherent states of two components $\left\vert \alpha
\right\rangle $ and $\left\vert -\alpha \right\rangle $ based on cavity QED
or circuit QED require a measurement on the states of auxiliary qubits or
qutrits [63,85-93].

\textbf{Possible experimental implementation.} Superconducting qubits play
important roles in quantum information processing [73,75,76,94-96]. In
addition, circuit QED is a realization of the physics of cavity QED with
superconducting qubits or other solid-state devices coupled to a microwave
cavity on a chip and has been considered as one of the most promising
candidates for quantum information processing [75,76,94-99]. Above, we
considered a general type of qubit for both the intracavity qubits and the
coupler qubit. As an example of experimental implementation, let us now
consider each qubit as a superconducting transmon qubit.

The dynamics of the lossy system, with finite qubit relaxation and dephasing
and photon lifetime included, is determined by the following master equation
\begin{eqnarray}
\frac{d\rho }{dt} &=&-i\left[ H_{Ik},\rho \right] +\sum_{j=1}^{3}\kappa _{j}%
\mathcal{L}\left[ a_{j}\right] +\sum_{j=1}^{3}\kappa _{j}^{\prime }\mathcal{L%
}\left[ b_{j}\right]  \nonumber \\
&&+\sum_{j=1}^{3}\left\{ \gamma _{j}\mathcal{L}\left[ \sigma _{j}^{-}\right]
\right\} +\sum_{j=1}^{3}\gamma _{j,\varphi }\left( \sigma _{z_{j}}\rho
\sigma _{z_{j}}-\rho \right)  \nonumber \\
&&+\gamma _{A}\mathcal{L}\left[ \sigma _{A}^{-}\right] +\gamma _{A,\varphi
}\left( \sigma _{z_{A}}\rho \sigma _{z_{A}}-\rho \right) ,
\end{eqnarray}%
where $H_{Ik}$ is either $H_{I1},$ $H_{I2},$ $H_{I3},$ or $H_{I4}$; $\ j$
represents qubit $j$ ($j=1,2,3$); $\sigma _{z_{j}}=\left\vert e\right\rangle
_{j}\left\langle e\right\vert -\left\vert g\right\rangle _{j}\left\langle
g\right\vert ,$ $\sigma _{z_{A}}=\left\vert e\right\rangle _{A}\left\langle
e\right\vert -\left\vert g\right\rangle _{A}\left\langle g\right\vert ;$ and
$\mathcal{L}\left[ \Lambda \right] =\Lambda \rho \Lambda ^{+}-\Lambda
^{+}\Lambda \rho /2-\rho \Lambda ^{+}\Lambda /2$, with $\Lambda
=a_{j},b_{j},\sigma _{j}^{-},\sigma _{A}^{-}$. In addition, $\kappa _{j}$ is
the decay rate of cavity $j$, $\kappa _{j}^{\prime }$ is that of NVE $j$ $,$
$\gamma _{j}~(\gamma _{A})$ is the energy relaxation rate of the level $%
\left\vert e\right\rangle $ of qubit $j~(A)$ , and $\gamma _{j,\varphi
}~(\gamma _{A,\varphi})$ is the dephasing rate of the level $\left\vert
e\right\rangle $ of qubit $j~(A)$.

The fidelity of the operation is given by [100]
\begin{equation}
\mathcal{F}=\sqrt{\left\langle \psi _{\mathrm{id}}\right\vert \rho
\left\vert \psi _{\mathrm{id}}\right\rangle },
\end{equation}%
where $\left\vert \psi _{\mathrm{id}}\right\rangle $ is the output state of
an ideal system (i.e., without dissipation and dephasing), while $\rho $ is
the output-state density operator of the system when the operations are
performed in a realistic physical system.

We now numerically calculate the fidelity of operation. Since the first
three steps employ resonant interactions, we will look at the operational
fidelity for each of these steps to see how short one should make the
typical operation time for each step to combat decoherence while still being
able to generate the entanglement with high fidelity. For simplicity, we
will consider the ideal output state of the previous step of operation as
the input state of the next step of operation when we analyze the
operational fidelities for the first three steps. In addition, we will
investigate the fidelity for the entire operation, which will be calculated
by numerically solving the master equation with the initial state of the
whole system as an input, but without making any approximation. Without loss
of generality and for simplicity, we will consider identical transmon
qubits, cavities, and NVEs. In this case, we have $g_{Aj}=g_{A},$ $%
g_{rj}\equiv g_{r},$ $g_{j}\equiv g,$ and $g_{b_{j}}\equiv g_{b}$ ($j=1,2,3$%
). We set $\Omega _{eg_{j}}=\Omega _{eg}$ and $\Omega _{j}=\Omega $ ($j=1,2,3
$). The decoherence times of transmon qubits and NVEs used in the numerical
simulation are: $\gamma _{j,\varphi }^{-1}=\gamma _{A,\varphi }^{-1}=15$ $%
\mu $s, $\gamma _{j}^{-1}=\gamma _{A}^{-1}=25$ $\mu $s, and $\kappa
_{j}^{\prime -1}=1$ ms (which is a conservative estimate compared with those
reported in experiments [53,101-103]). In addition, we choose $\kappa
_{j}^{-1}=1$ $\mu $s in the numerical simulation ($j=1,2,3$).

\textbf{A. Fidelity for the first three steps.} The operation fidelities are
plotted in Figs.~3(a,b,c), which are for step 1, step 2, and step 3,
respectively. Figure~3 shows that the fidelity for step 1, step 2, or step 3
increases drastically with $g_{A},$ $g_{r}$, or $\Omega _{eg}$ and reaches a
high value $\ 0.998\leq \mathcal{F}\leq 1$ for $g_{A}/\left( 2\pi \right)
,g_{A}/\left( 2\pi \right) ,\Omega _{eg}/\left( 2\pi \right) \in \lbrack 5$
MHz$,50$ MHz$]$, which corresponds to the operation time $\sim 3$--$30$ ns.
The analysis given here demonstrates that in order to combat decoherence
while obtain the entanglement with a high fidelity $\sim 1$, one should make
the typical operation time within a few nanoseconds for each of the first
three steps, and a high fidelity $\geq $ $0.998$ can be achieved even by
increasing the operation time to $\sim 30$ ns.

\textbf{B. Fidelity for the entire operation.} The fidelity for the entire
operation is calculated based on Eq.~(27), where the ideal output state is $%
\left\vert \psi _{\mathrm{id}}\right\rangle =\left\vert \varphi
\right\rangle \otimes \prod_{j=1}^{3}\left\vert 0\right\rangle
_{c_{j}}\left\vert g\right\rangle _{A}$ [with $\left\vert \varphi
\right\rangle $ given by Eq.~(20) or Eq.~(23)] and $\rho $ is obtained by
numerically solving the master equation (26) for an initial input state $%
\left\vert \psi _{\mathrm{in}}\right\rangle =\prod\limits_{j=1}^{3}|g\rangle
_{j}\prod\limits_{j=1}^{3}|0\rangle _{b_{j}}\prod\limits_{j=1}^{3}|0\rangle
_{c_{j}}\otimes |e\rangle _{A}.$ We choose $g_{A}/\left( 2\pi \right) =50$
MHz, $g_{r}/\left( 2\pi \right) =g/\left( 2\pi \right) =5$ MHz, and $%
g_{b}/\left( 2\pi \right) \sim 4$ MHz [44]. We here select $g_{r}=g$ because
the resonant coupling constant $g_{r}$ and the off-resonant coupling
constant $g$ are both the same order of magnitude for superconducting
qubits. Other parameters used in the numerical simulation are: $\Omega
_{eg}/\left( 2\pi \right) =50$ MHz, $\Omega /\left( 2\pi \right) =100$ MHz
(available in experiments [104,105]), and $\delta _{a_{j}}=7.2g_{j}$
(obtained by numerically optimizing the system parameters). With the choice
of these parameters, the fidelity versus $D=\delta _{b_{j}}/g_{b_{j}}$ is
plotted in Fig.~4, which demonstrates that for $D\sim $ $9,$ a high fidelity
$\sim 93.2\%$ can be achieved for the state $\left\vert \varphi
\right\rangle $ with $\left\vert \beta \right\vert =1.2$. For $D\sim $ $9$,
the entire operation time is estimated to be $\sim 1.14$ $\mu $s, much
shorter than the decoherence times of transmon qubits and NVEs used in our
numerical simulation but a little longer than the cavity decay time.
Figure~4 also shows that the fidelity heavily depends on $D$ (or the
detuning $\delta _{b_{j}}$). The fidelity reaches its maximum as $D$
increases to $9.$ However, it drops down when $D$ becomes larger than $9$.
This means that further increasing the detuning $\delta _{b_{j}}$ will have
an adverse effect on the fidelity. The interpretation for this is: As the
detuning $\delta _{b_{j}}$ becomes larger than the optimum value $9g_{b_{j}}$
($2\pi \times 36$ MHz) (i.e., the value where the large detuning is well
satisfied), the NVE-cavity coupling becomes weaker, which increases the
operation time and thus the effect of decoherence from transmon qubits and
NVEs on the fidelity becomes more apparent.

Note that although the entire operation time is longer than the cavity decay
time used in our numerical simulations, the effect of the cavity decay on
the fidelity is negligible. This is because: the first three steps are
completed within a very short time due to using the resonant interaction,
and (as illustrated in Fig.~5) the number of photons occupied in each cavity
during the last step of operation is quite low due to using a large-detuning
technique. Indeed, to reduce decoherence from the cavity decay, one can
employ a longer cavity-decay time in the numerical simulation, which however
would require cavities with a higher-$Q$ quality factor and thus may pose a
challenge in experiments.

Figure~5 is plotted by choosing the detuning $D=9$ and using the same
parameters for Fig.~4. For simplicity, Fig.~5 only shows the curves
corresponding to the operation time $t-t_{0}$ required for the last step of
operation. Here, $t$ is the entire operation time while $t_{0}$ is the time
required for the first three steps of operation. For the values of $g_{A},$ $%
g_{r},$ and $\Omega _{eg}$ chosen above, $t_{0}$ is $\sim 36$ ns. The blue
curve represents the fidelity, which is calculated for an ideal state $%
\left\vert \psi _{\mathrm{id}}\right\rangle $ ($\left\vert \varphi
\right\rangle $) with $\left\vert \beta \right\vert =1.2$. The red curve
represents the value of $\left\vert \beta \right\vert /2$ or $\left\vert
-\beta \right\vert /2$. The green curve indicates the average photon number
for each cavity. Figure~5 indicates that the fidelity increases when $%
t-t_{0} $ approaches $1.08$ $\mu $s (which is the time required for the last
step of operation for preparing the desired state $\left\vert \varphi
\right\rangle $ with $\left\vert \beta \right\vert =1.2$). The maximum
fidelity depicted in Fig.~5 is in good agreement with that shown in Fig.~4
for $D=9$. In addition, the green curve shows that the average number of
photons excited in each cavity is less than 0.02, implying that the cavity
photons are almost not excited during the last step of operation.

According to experimental reports [81,82], the cavity frequency can be
rapidly adjusted by $\Delta \omega _{c}/\left( 2\pi \right) =500\sim 740$
MHz. As a conservative consideration, for $\Delta \omega _{c}/\left( 2\pi
\right) =500$ MHz, the detuning $\delta _{b_{j}}$ changes to $\widetilde{%
\delta }_{b_{j}}=9g_{b_{j}}+2\pi \times 500$ MHz, which can be further
written as $\widetilde{\delta }_{b_{j}}/g_{b_{j}}\sim 134$ for $%
g_{bj}/\left( 2\pi \right) \equiv g_{b}/\left( 2\pi \right) =4$ MHz chosen
above. This result shows that the decoupling of the cavities with the NVEs,
which was required during the $W$-state preparation, can be well met by
adjusting the cavity frequency. As discussed previously, the coupling or
decoupling of the qubits with the cavities can be readily made by adjusting
the level spacings of the qubits.

$T_{1}$ (energy relaxation time) and $T_{2}$ (dephasing time) can be made to
be on the order of $20$--$80$ $\mu $s for state-of-the-art superconducting
transmon devices [101-103]. In addition, the lifetime of an NVE can reach $%
\sim 1.2$ s according to recent experimental reports [53]. The typical
transition frequency of a transmon qubit is between 2 and 10 GHz [77,106].
As an example, consider each cavity of frequency $\nu _{c}\sim 5$ GHz.
Hence, for the $\kappa _{j}^{-1}$ used in the numerical calculation, the
required quality factor of each cavity is $Q_{j}\sim 3.1\times 10^{4},$
which is accessible in experiments because a quality factor $Q\sim 5\times
10^{4}$ for CPW resonators with loaded NVEs has been experimentally
demonstrated [44]. The analysis given here shows that a high-fidelity
implementation of the three-NVE $W$-type entangled coherent state $%
\left\vert W_{1}\right\rangle $, $\left\vert W_{2}\right\rangle $, $%
\left\vert W_{3}\right\rangle $, or $\left\vert W_{4}\right\rangle $
described by Eq.~(24) is feasible within present-day circuit QED techniques.

\textbf{Quantum state transfer.} Consider a cavity and an NVE inside the
cavity. Based on Eq.~(35) (see the Methods section), the NVE-cavity
interaction Hamiltonian can be written as
\begin{equation}
H_{I}=g_{b}(a^{\dagger }b+ab^{\dagger }),
\end{equation}%
where we set $\delta =\omega _{b}-\omega _{c}=0$. Assume now that the
initial state of the cavity and the NVE is given by $|0\rangle _{c}\otimes
|\beta \rangle _{\mathrm{NVE}}$, where $|0\rangle _{c}$ is the vacuum state
of the cavity while $|\beta \rangle _{\mathrm{NVE}}$ is the coherent state
of the NVE, given by $|\beta \rangle _{\mathrm{NVE}}=\exp (-\frac{1}{2}%
|\beta |^{2})\sum\limits_{n=0}^{\infty }\frac{\beta ^{n}}{\sqrt{n!}}%
|n\rangle _{\mathrm{NVE}}$. In terms of $|n\rangle _{\mathrm{NVE}}=\frac{%
b^{\dagger n}}{\sqrt{n!}}|0\rangle _{\mathrm{NVE}}$, one can describe the
system initial state as
\begin{equation}
|0\rangle _{c}\otimes |\beta \rangle _{\mathrm{NVE}}=\exp (-|\beta
|^{2}/2)\sum\limits_{n=0}^{\infty }\frac{\beta ^{n}\left( b^{\dagger
}\right) ^{n}}{n!}|0\rangle _{\mathrm{NVE}}|0\rangle _{c}.
\end{equation}

Making use of the Hamiltonian (28), we can obtain the transformations $%
e^{-iH_{I}t}b^{\dagger }e^{iH_{I}t}=\cos (g_{b}t)b^{\dagger }+i\sin
(g_{b}t)a^{\dagger }$. For $g_{b}t=\pi /2$, one has $e^{-iH_{I}t}b^{\dagger
}e^{iH_{I}t}=ia^{\dagger }$. Under the Hamiltonian (28) and after an
evolution time $t=\pi /(2g_{b})$, the state of the system can be written as
\begin{eqnarray}
&&e^{-iH_{I}t}|0\rangle _{c}\otimes |\beta \rangle _{\mathrm{NVE}}  \nonumber
\\
&=&e^{-iH_{I}t}\exp (-|\beta |^{2}/2)\sum\limits_{n=0}^{\infty }\frac{\beta
^{n}(b^{\dagger })^{n}}{n!}|0\rangle _{c}|0\rangle _{\mathrm{NVE}}  \nonumber
\\
&=&e^{-iH_{I}t}\exp (-|\beta |^{2}/2)\sum\limits_{n=0}^{\infty }\frac{\beta
^{n}(b^{\dagger })^{n}}{n!}e^{iH_{I}t}e^{-iH_{I}t}|0\rangle _{c}|0\rangle _{%
\mathrm{NVE}}  \nonumber \\
&=&\exp (-|\beta |^{2}/2)\sum\limits_{n=0}^{\infty }\frac{\beta ^{n}}{n!}%
e^{-iH_{I}t}(b^{\dagger })^{n}e^{iH_{I}t}|0\rangle _{c}|0\rangle _{\mathrm{%
NVE}}  \nonumber \\
&=&\exp (-|\beta |^{2}/2)\sum\limits_{n=0}^{\infty }\frac{(i\beta )^{n}}{n!}%
(a^{\dagger })^{n}|0\rangle _{c}|0\rangle _{\mathrm{NVE}}  \nonumber \\
&=&|i\beta \rangle _{c}\otimes |0\rangle _{\mathrm{NVE}},
\end{eqnarray}%
where we have used $e^{-iH_{I}t}(b^{\dagger })^{n}e^{iH_{I}t}=(ia^{\dagger
})^{n}$ and $e^{-iH_{I}t}|0\rangle _{c}|0\rangle _{\mathrm{NVE}}=|0\rangle
_{c}|0\rangle _{\mathrm{NVE}}$.

In the same manner, after an evolution time $t=\pi /2g_{b}$, the state $%
|0\rangle _{c}|-\beta \rangle _{\mathrm{NVE}}$ of the cavity and the NVE is
transformed to $|-i\beta \rangle _{c}\otimes |0\rangle _{\mathrm{NVE}}$.
Given the above results, one can transfer a macroscopic $W$-type entangled
coherent state from the NVEs into the cavities. For instance, the above
state $\left\vert W_{1}\right\rangle $ of the three NVEs is transferred onto
the three cavities, becoming
\begin{equation}
|W_{1}\rangle _{c}=\frac{1}{\sqrt{3}}\left( |-i\beta \rangle |i\beta \rangle
|i\beta \rangle +|i\beta \rangle |-i\beta \rangle |i\beta \rangle +|i\beta
\rangle |i\beta \rangle |-i\beta \rangle \right) .
\end{equation}

\section*{Discussion}

A method has been presented to generate a \textit{continuous-variable} $W$%
-type entangled coherent state of NVEs in circuit QED. As shown above, this
proposal offers some distinguishing features and advantages: (i) The $W$
state is prepared in the NVEs (quantum memories), while not prepared with
the cavity photons. (ii) Because of NVE's long decoherence time, the
prepared $W$ state can be stored in the NVEs for a long time, when compared
with storing it via cavity photons. (iii) For most of the operation time,
cavity photons are virtually excited, and thus decoherence caused by the
cavity decay is significantly suppressed. (iv) Because each cavity remains
in a vacuum state after the state preparation, the decoherence due to the
cavity decay is avoided during storing the prepared $W$ state via the NVEs.
(v) The state preparation does not require that each NVE is initially
prepared in a coherent state, which should significantly reduce its
experimental difficulty. (vi) Moreover, the proposal employs only one
external-cavity coupler qubit. The prepared $W$ state of NVEs can be mapped
onto the cavities by local operations. This proposal is quite general and
can be extended to create the proposed $W$ state with atomic ensembles or
other spin ensembles distributed over different cavities. Our numerical
simulations show that the high-fidelity implementation of $W$-type entangled
coherent states with three NVEs is feasible with current circuit QED
technology.

\section*{Methods}

\textbf{NVE-cavity interaction Hamiltonian. }As shown in Fig.~6(a), the
energy levels of an NV center consist of a ground state $^{3}A$, an excited
state $^{3}E$ and a metastable state $^{1}A$. Both $^{3}A$ and $^{3}E$ are
spin triplet states while the metastable $^{1}A$ is a spin singlet state
[107,108]. The NV center has an $S=1$ ground state with zero-field splitting
$D_{gs}/\left( 2\pi \right) =2.88$ GHz between the $|m_{s}=0\rangle $ and $%
|m_{s}=\pm 1\rangle $ levels [Fig.~6(a)]. By applying an external magnetic
field along the crystalline axis of the NV center [47,$83$], an additional
Zeeman splitting between $|m_{s}=\pm 1\rangle $ sublevels occurs [Fig.~6(b)].

If we need to eliminate the coupling of the cavity with the NV center, one
can adjust the cavity frequency $\omega _{c}$ to have $\omega _{c}$
sufficiently larger than $\omega _{0,+1}$ and $\omega _{0,-1},$ such that
the cavity mode is highly detuned (decoupled) from both the $|m_{s}=0\rangle
\leftrightarrow |m_{s}=-1\rangle $ transition and the $|m_{s}=0\rangle
\leftrightarrow |m_{s}=+1\rangle $ transition [Fig.~6(c)]$.$ Here, $\omega
_{0,+1}$ ($\omega _{0,-1}$) is the transition frequency between the two
levels $|m_{s}=0\rangle $ and $|m_{s}=+1\rangle $ ($|m_{s}=-1\rangle $)$.$
On the other hand, one can adjust the cavity frequency such that the cavity
mode is coupled with the transition between the ground level $%
|m_{s}=0\rangle $ and the excited level $|m_{s}=+1\rangle $, but still
decoupled from the transition between the two levels $|m_{s}=0\rangle $ and $%
|m_{s}=-1\rangle $ [Fig.~6(d)]. Note that for a superconducting transmission
line resonator, the rapid tuning of cavity frequencies by a few hundred MHz
in $1$--$2$ nanoseconds has been demonstrated in experiments [81,82]).
During the $W$-state preparation described in the Results section, we assume
that the level splitting of the NV center is fixed.

An NV center is usually treated as a spin while an ensemble of NV centers is
treated as a spin ensemble (i.e., an NVE). Let an NVE be placed at an
antinode of a single mode of the electromagnetic field. When the cavity is
coupled to the $|m_{s}=0\rangle \leftrightarrow |m_{s}=+1\rangle $
transition, but decoupled from the $|m_{s}=0\rangle \leftrightarrow
|m_{s}=-1\rangle $ transition [Fig.~6(d)], the system Hamiltonian in the
interaction picture reads (in units of $\hbar =1$)
\begin{equation}
H_{C,\mathrm{NVE}}=\sum\limits_{k=1}^{N}g_{k}(a^{\dagger }\tau
_{k}^{-}e^{i\delta t}+a\tau _{k}^{+}e^{-i\delta t}),
\end{equation}%
where $\delta =\omega _{c}-\omega _{0,+1}$, $\omega _{c}$ is the
eigenfrequency of the cavity mode, $a$ $(a^{\dagger })$ is the corresponding
annihilation (creation) operator of the cavity mode, $\tau
_{k}^{+}=|m_{s}=+1\rangle _{k}\langle m_{s}=0|$ and $\tau
_{k}^{-}=|m_{s}=0\rangle _{k}\langle m_{s}=+1|$ are the raising and lowering
operators for the $k$th spin, and $g_{k}$ is the coupling strength between
the cavity and the $k$th spin. We then define a collective operator
\begin{equation}
b^{\dagger }=\left( \frac{1}{\sqrt{N}}\right) \left( \frac{1}{\overline{g}}%
\right) \sum\limits_{k=1}^{N}g_{k}\tau _{k}^{+},
\end{equation}%
with $\overline{g}^{2}=\sum\limits_{k=1}^{N}|g_{k}|^{2}/N$, and $\overline{g}
$ is the root mean square of the individual couplings.

Under the condition of a large $N$ and a very small number of excited spins
(compared to the number $N$), $b^{\dagger }$ behaves as a bosonic operator
and the spin ensemble behaves as a bosonic mode. Thus, we have $%
[b,b^{\dagger }]\approx 1,$ and $b^{\dagger }b|n\rangle _{b}=n|n\rangle _{b}$
[48,109], where
\begin{equation}
|n\rangle _{b}=\frac{1}{\sqrt{n!}}(b^{\dagger })^{n}|0\rangle _{b}
\end{equation}%
with $|0\rangle _{b}=|m_{s}=0\rangle _{1}|m_{s}=0\rangle _{2}\cdots
|m_{s}=0\rangle _{N}$. It is easy to verify that the frequency $\omega _{b}$
of the bosonic mode describing the NVE is equal to the transition frequency $%
\omega _{0,+1}$ between the ground level $|m_{s}=0\rangle $ and the excited
level $|m_{s}=+1\rangle $ of each spin (i.e. $\omega _{b}=\omega _{0,+1}$).
For simplicity we have defined $|m_{s}=+1\rangle =|+1\rangle $ and $%
|m_{s}=0\rangle =|0\rangle $.

Therefore, the Hamiltonian (32) can be further rewritten as
\begin{equation}
H_{C,\mathrm{NVE}}=g_{b}(e^{i\delta t}a^{\dagger }b+e^{-i\delta
t}ab^{\dagger }),
\end{equation}%
with $g_{b}=\sqrt{N}\overline{g}$. Based on Eq.~(35), one can find that for
the case of three NVEs each placed in a cavity, the Hamiltonian for the
three NVEs interacting with their respective cavities would be the second
term of Eq.~(11).

\textbf{NVE-cavity coupling selection.} During the last step of the $W$
state preparation, we would require the coupling of each cavity with the $%
|m_{s}=0\rangle \leftrightarrow |m_{s}=+1\rangle $ transition while
decoupling each cavity from the $|m_{s}=0\rangle \leftrightarrow
|m_{s}=-1\rangle $ transition. The advantage of this is that the created $W$
state has a mode frequency equal to $\omega _{0,+1}$, which is adjustable by
varying the magnetic field applied to the NVEs [Fig.~6(c), Fig.~6(d)].
Instead of using the coupling of each cavity with the $|m_{s}=0\rangle
\leftrightarrow |m_{s}=+1\rangle $ transition, one can employ the coupling
of each cavity with the $|m_{s}=0\rangle \leftrightarrow |m_{s}=\pm 1\rangle
$ transition (i.e., the transition between the ground state $|m_{s}=0\rangle
$ and the degenerate excited states $|m_{s}=\pm 1\rangle ).$ However, there
is an inevitable shortcoming, i.e., the created $W$ state has a fixed mode
frequency, which is equal to $\omega _{0,\pm 1}=2\pi \times 2.88$ GHz
[Fig.~6(a)] and thus cannot be adjusted.

\begin{addendum}

\item[Acknowledgments]

We very gratefully acknowledge Dr.~Anton Frisk Kockum for a critical
reading of the manuscript. C.P. Yang was supported in part by the National
Natural Science Foundation of China under Grant Nos.~11074062 and 11374083,
the Zhejiang Natural Science Foundation under Grant No.~LZ13A040002, the
funds from Hangzhou Normal University under Grant Nos.~HSQK0081 and
PD13002004, and the funds from Hangzhou City for the Hangzhou-City Quantum
Information and Quantum Optics Innovation Research Team. J.M. Liu was
supported in part by the National Natural Science Foundation of China under
Grant Nos. 11174081, 11034002, and 11134003, and the National Basic Research
Program of China under Grant Nos. 2011CB921602 and 2012CB821302. F. Nori was
partially supported by the RIKEN iTHES Project, the MURI Center for Dynamic Magneto-Optics
via the AFOSR award number FA9550-14-1-0040, and a Grant-in-Aid for Scientific Research (A).

\item[Author contributions]
T. L, S. J. X. and C. P. Y conceived the idea. Q. P. S carried out all calculations under the guidance of C.P.Y. and F.N. All the authors discussed the results. T. L, J. M. L, C. P. Y, and F. N. contributed to the writing of the manuscript.

\item[Additional information]
Competing financial interests: The authors declare no competing financial
interests.
\end{addendum}

\clearpage

\textbf{Figure 1:} (a) Setup of the hybrid system consisting of a coupler
qubit $A$ and three cavities each hosting a qubit (a dark dot) and a
nitrogen-vacancy center ensemble (a green oval). $C_{1}$, $C_{2}$ and $C_{3}$
represent capacitors. An intracavity qubit can be an atom or a solid-state
qubit. The coupler qubit $A$ can be a quantum dot or a superconducting
qubit. (b) Illustration of the decoupling among qubit $A,$ cavity $j,$ NVE $%
j $ and qubit $j$ ($j=1,2,3$) before the $W$-state preparation. (c) The
resonant interaction between qubit $A$ and cavity $j$ with coupling constant
$g_{A_{j}}$ (used in step 1). (d) The resonant interaction between qubit $j$
and cavity $j$ with resonant coupling constant $g_{rj}$ (used in step 2).
(e) The resonant interaction between qubit $j$ and the pulse with Rabi
frequency $\Omega _{eg_{j}}$ (applied for step 3). (f) The dispersive
interaction between cavity $j$ and qubit $j$ with coupling constant $g_{j}$
and detuning $\delta _{a_{j}}$, the dispersive interaction between cavity $j$
and NVE $j$ with coupling constant $g_{b_{j}}$ and detuning $\delta _{b_{j}}$%
, as well as the resonant interaction between qubit $j$ and the pulse with
Rabi frequency $\Omega_j$ (applied for step 4). Here, $\delta
_{a_{j}}=\omega _{c_{j}}-\omega _{eg_{j}}$, with the transition frequency $%
\omega _{eg_{j}}$ of qubit $j$ and the frequency $\omega _{c_{j}}$ of cavity
$j$, $\delta _{b_{j}}=\omega _{c_{j}}-\omega _{b_{j}}$ and $\delta
_{c_{j}}=\delta _{a_{j}}-\delta _{b_{j}}$, with $\omega _{b_{j}}$ being the
frequency of a bosonic mode describing NVE $j$. Since qubit $A$ is not
involved during the operation of step 4, qubit $A$ is dropped off in (f) for
simplicity. Note that in (b) and (e), the frequency of cavity $j$ is highly
detuned from those of qubit $A,$ qubit $j$ and NVE $j,$ while in (f) the
frequeny of cavity $j$ is adjusted such that cavity $j$ is dispersively
coupled to qubit $j$ and NVE $j.$ The bottom dark solid line in (b)-(f) also
represents the ground state (i.e., the vacuum state) of cavity $j$.

\bigskip \textbf{Figure 2:} Diagram of a coupler qubit $A$ and $n$ cavities
each hosting a qubit (a dark dot) and a NVE (a green Oval). Qubit $A$ is
capacitively coupled to each cavity.

\textbf{Figure 3:} (a) Fidelity for step 1. (b) Fidelity for step 2. (c)
Fidelity for step 3. \bigskip

\textbf{Figure 4:} Fidelity $\mathcal{F}$ versus reduced detuning $D=\delta
_{b_j}/g_{b_j}$. The red squares correspond to the case without considering
the errors and decoherence for the first three-step operation, while the
blue dots correspond to the case after the errors and decoherence for the
first three-step operation are taken into account. The parameters used here
are described in the text. \bigskip

\textbf{Figure 5:} The operational fidelity $\mathcal{F}$, the amplitude $%
\left| \beta \right| $ (or $\left| -\beta \right| $), and the photon number
of each cavity versus $t-t_0$ (i.e., the time required for the last step of
operation). The blue curve represents the operational fidelity, which is
calculated for an ideal state $\left| \psi _{id}\right\rangle $ ($\left|
\varphi \right\rangle $) with $\left| \beta \right| =1.2$. The red curve
represents the value of $\left| \beta \right|/2 $ or $\left| -\beta
\right|/2 $. The green curve indicates the photon number (enlarged 10 times)
of each cavity. For $t-t_0=1.08$ $\mu $s (i.e., the time required for
preparing the state $\left| \varphi \right\rangle$ with $|\beta|=1.2$ during
the last step of operation), the fidelity $\mathcal{F}$ reaches the maximum.
All curves are plotted for reduced detuning $D=\delta _{b_j}/g_{b_j}=9$ and
parameters used in Fig.~4. \bigskip

\textbf{Figure 6:} (a) Schematic diagram of electronic and spin energy
levels of a nitrogen-vacancy center. (b) The ground electronic-spin levels
of an NV center in the presence of an external magnetic field parallel to
the crystalline axis. Here $B$ and $E$ represent the magnetic field and
energy, respectively. (c) Illustration of the cavity decoupled from the NV
center. Here, $\omega_c$ is the cavity frequency, while $\omega_{0,-1}$ ($%
\omega_{0,+1}$) is the energy gap between the $|m_s=0\rangle $ and $%
|m_s=-1\rangle $ ($|m_s=+1\rangle $) levels of the NV center. The cavity
frequency $\omega _{c}$ is sufficiently larger than $\omega _{0,+1}$ and $%
\omega _{0,-1}$, such that the cavity mode is highly detuned (decoupled)
from both the $|m_{s}=0\rangle\leftrightarrow |m_{s}=-1\rangle $ transition
and the $|m_{s}=0\rangle \leftrightarrow |m_{s}=+1\rangle $ transition. (d)
Illustration of the cavity being coupled to the $|m_s=0\rangle
\leftrightarrow |m_s=+1\rangle $ transition with a detuning $\delta =\omega
_c-\omega _{0,+1}$, but decoupled from the $|m_s=0\rangle \leftrightarrow
|m_s=-1\rangle $ transition of the NV center. \bigskip

\clearpage

\begin{figure}[tbp]
\begin{center}
\epsfig{file=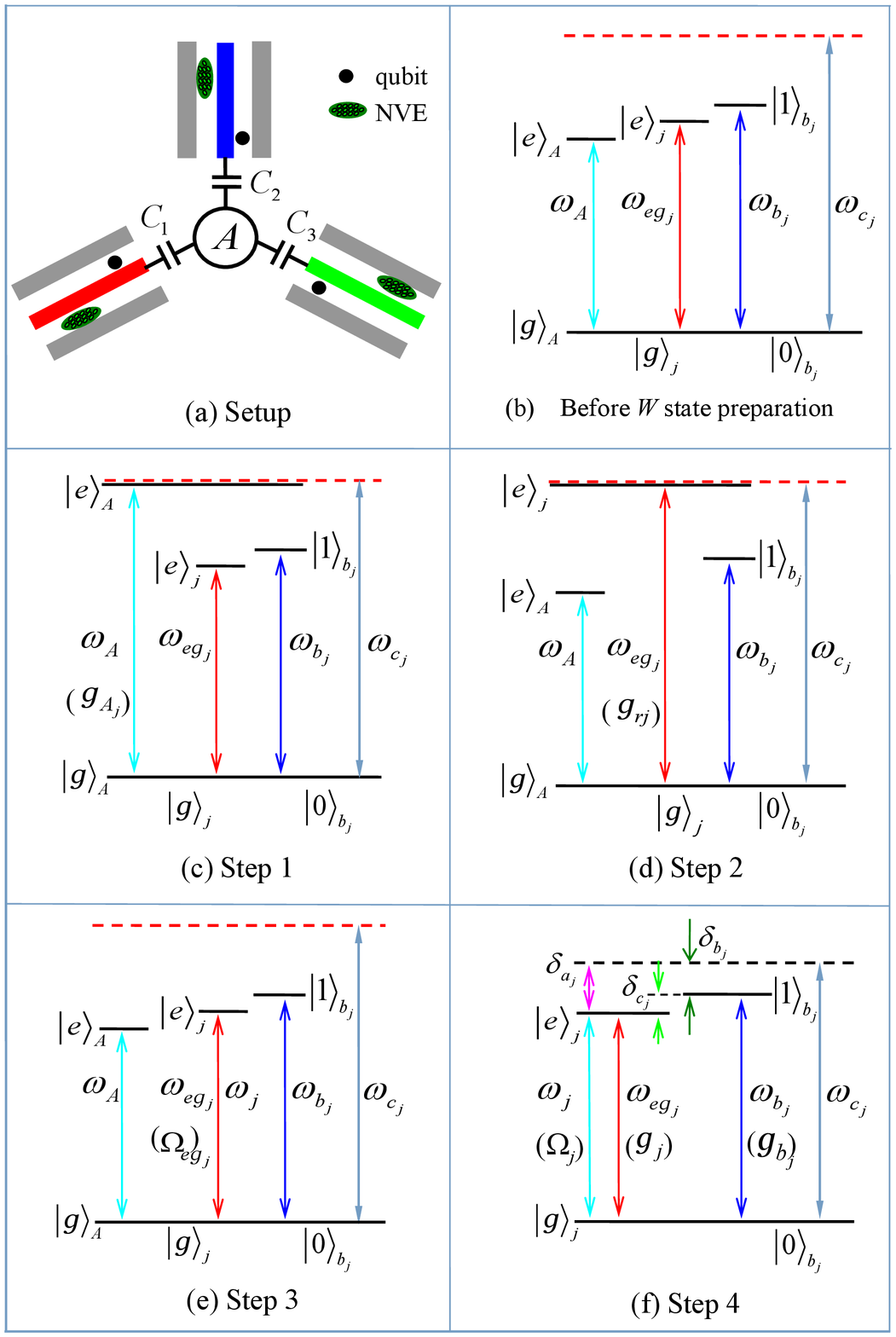,width=13cm}
\end{center}
\caption{}
\label{fig:1}
\end{figure}

\begin{figure}[tbp]
\begin{center}
\epsfig{file=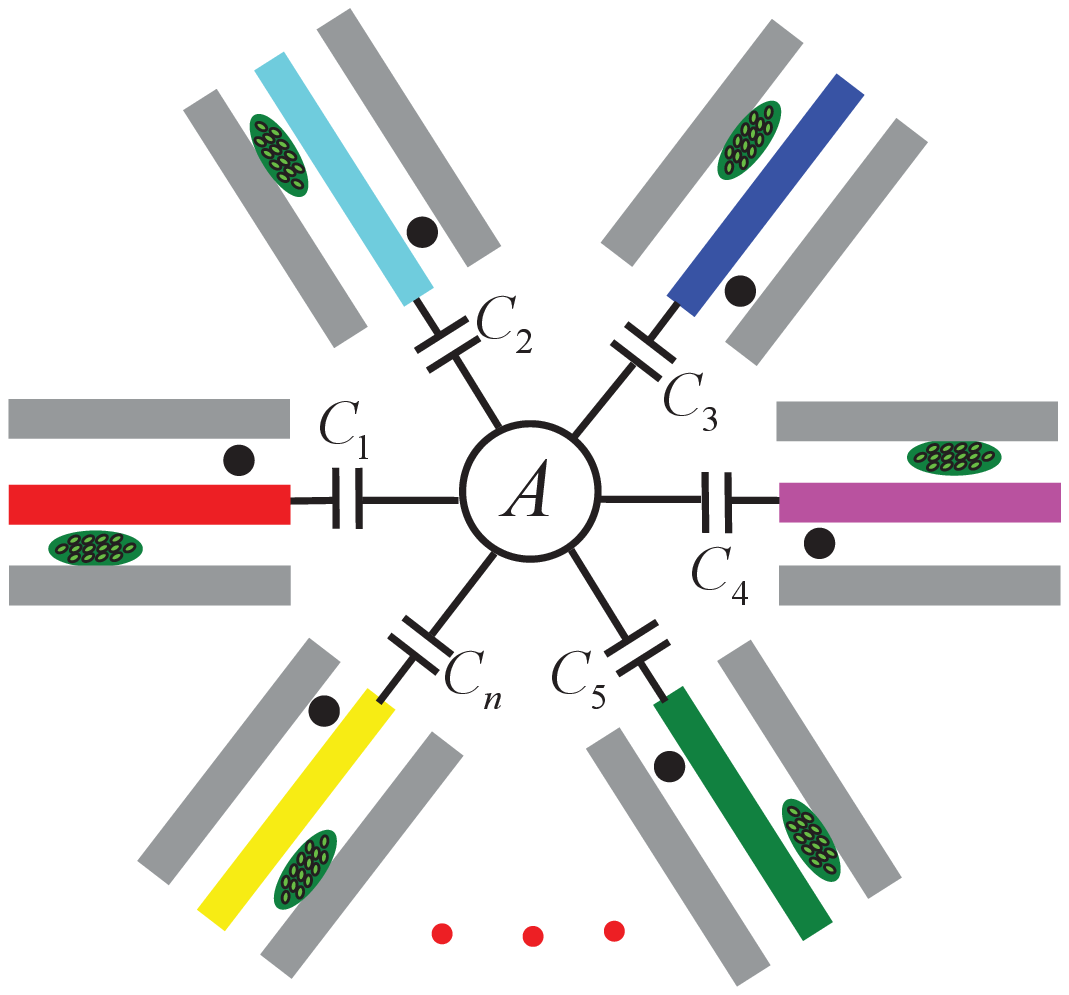,width=15.5cm}
\end{center}
\caption{}
\label{fig:2}
\end{figure}

\begin{figure}[tbp]
\begin{center}
\epsfig{file=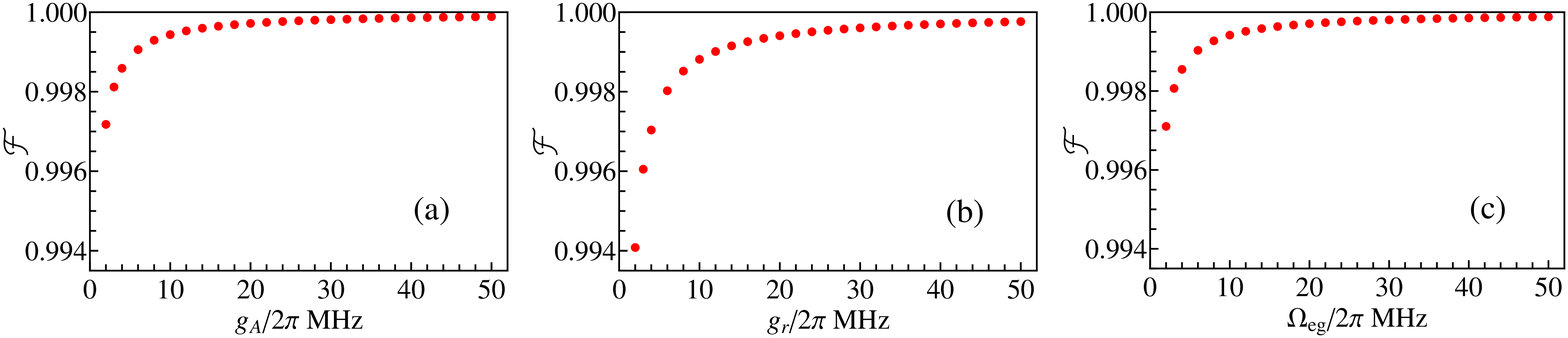,width=15.5cm}
\end{center}
\caption{}
\label{fig:3}
\end{figure}

\begin{figure}[tbp]
\begin{center}
\epsfig{file=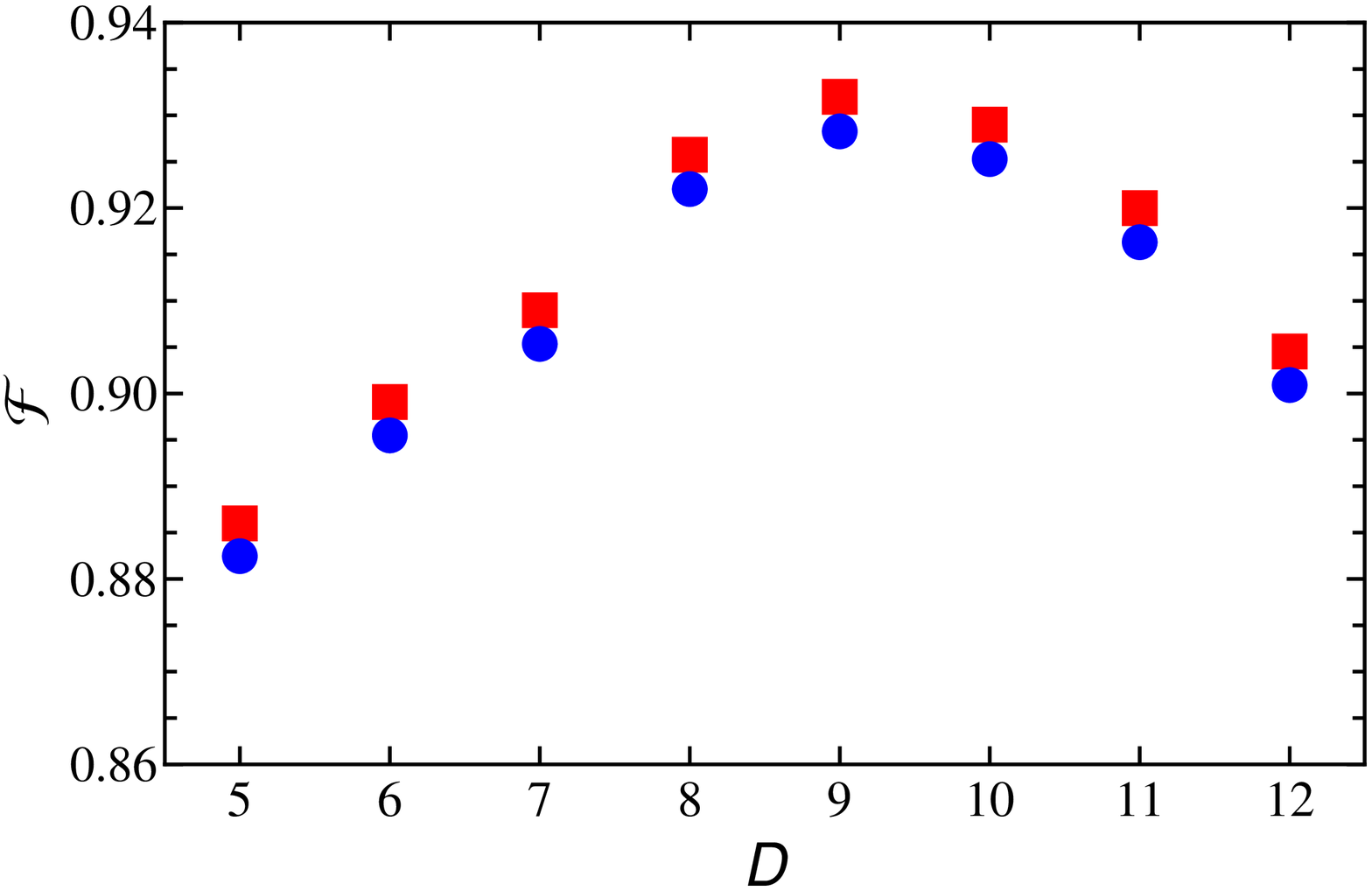,width=13cm}
\end{center}
\caption{}
\label{fig:4}
\end{figure}

\begin{figure}[tbp]
\begin{center}
\epsfig{file=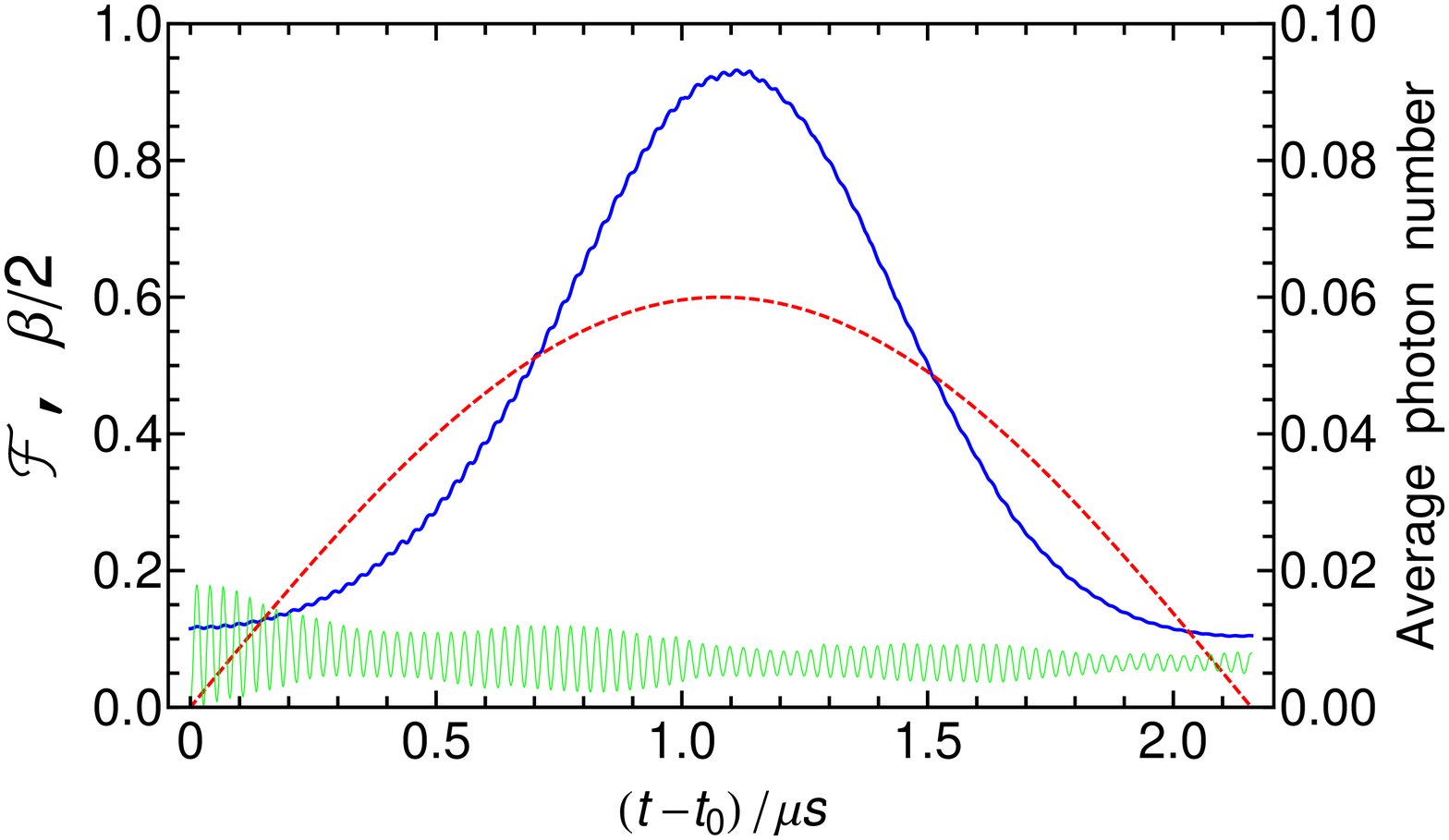,width=14cm}
\end{center}
\caption{}
\label{fig:5}
\end{figure}

\begin{figure}[tbp]
\begin{center}
\epsfig{file=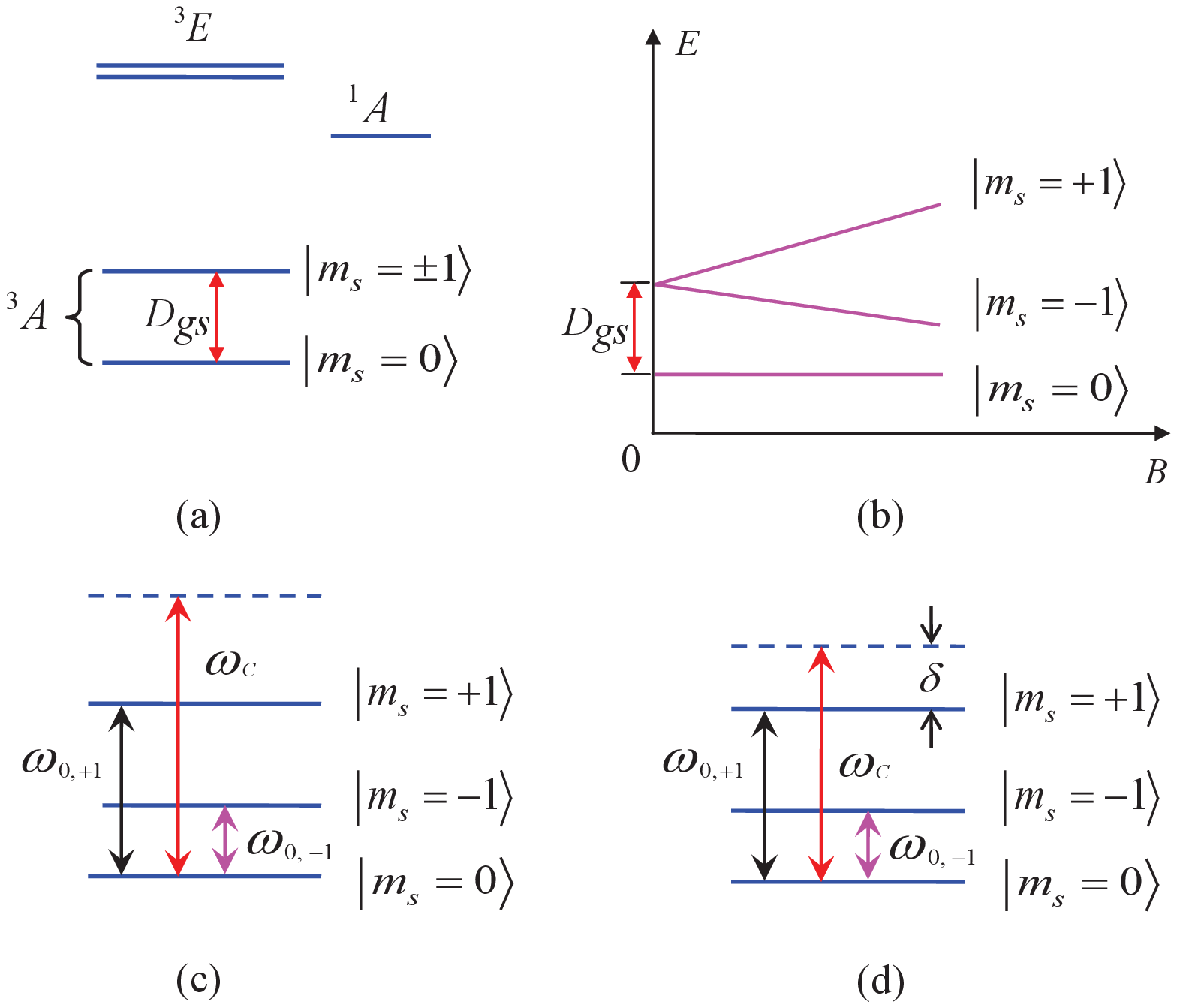,width=30cm}
\end{center}
\caption{}
\label{fig:6}
\end{figure}

%\begin{figure}
%\begin{center}
%\epsfig{file=Fig-S1.eps,width=12cm}
%\end{center}\caption{}
%\label{Schematic diagram of quantum multiple access network with
%chaotic phase shifters}
%\end{figure}

\end{document}